\documentclass[aps,prl,twocolumn,superscriptaddress,showpacs]{revtex4}

\usepackage{graphicx}

\begin{document}


\title{Two-dimensional resonant magnetic excitation in BaFe$_{1.84}$Co$_{0.16}$As$_2$}


\author{M. D. Lumsden}
\author{A. D. Christianson}
\affiliation{Oak Ridge National Laboratory, Oak Ridge, TN 37831, USA}
\author{D. Parshall}
\affiliation{Department of Physics and Astronomy, University of Tennessee, Knoxville, TN 37996}
\author{M. B. Stone}
\author{S. E. Nagler}
\author{G.J. MacDougall}
\author{H. A. Mook}
\affiliation{Oak Ridge National Laboratory, Oak Ridge, TN 37831, USA}
\author{K. Lokshin}
\affiliation{Department of Materials Science and Engineering, University of Tennessee, Knoxville, TN 37996}
\author{T. Egami}
\affiliation{Oak Ridge National Laboratory, Oak Ridge, TN 37831, USA}
\affiliation{Department of Physics and Astronomy, University of Tennessee, Knoxville, TN 37996}
\affiliation{Department of Materials Science and Engineering, University of Tennessee, Knoxville, TN 37996}
\author{D. L. Abernathy}
\affiliation{Oak Ridge National Laboratory, Oak Ridge, TN 37831, USA}
\author{E. A. Goremychkin}
\affiliation{Argonne National Laboratory, Argonne, IL 60439, USA}
\affiliation{ISIS Facility, Rutherford Appleton Laboratory, Chilton, Didcot OX11 0QX, United Kingdom}
\author{R. Osborn}
\affiliation{Argonne National Laboratory, Argonne, IL 60439, USA}
\author{M. A. McGuire}
\author{A. S. Sefat}
\author{R. Jin}
\author{B. C. Sales}
\author{D. Mandrus}
\affiliation{Oak Ridge National Laboratory, Oak Ridge, TN 37831, USA}

\begin{abstract}

Inelastic neutron scattering measurements on single crystals of superconducting
BaFe$_{1.84}$Co$_{0.16}$As$_2$ reveal a magnetic excitation located at wavevectors (1/2 1/2 L) in tetragonal notation. On cooling below $T_C$, a clear resonance peak is observed at this wavevector with an energy of 8.6(0.5) meV, corresponding to 4.5(0.3) k$_B$$T_C$.  This is in good agreement with the canonical value of 5 k$_B$$T_C$ observed in the cuprates. The spectrum shows strong dispersion in the tetragonal plane but very weak dispersion along the c-axis, indicating that the magnetic fluctuations are two-dimensional in nature.  This is in sharp contrast to the anisotropic three dimensional spin excitations seen in the undoped parent compounds.

\end{abstract}


\maketitle

Understanding the physics of superconductivity in high-$T_c$ cuprates and other unconventional superconductors remains a central unresolved problem at the forefront of condensed matter physics.  One widespread school of thought maintains that magnetic fluctuations are intimately involved in the pairing mechanism.  This view is supported by a growing number of neutron scattering investigations showing the appearance of a magnetic excitation coincident with the onset of superconductivity \cite{rossat-mignod,mook,fong,dai,he,stock,metoki,sato}.  The spectrum shows a resonance at a wavevector related to the antiferromagnetic order in the non-superconducting parent compounds.  The apparent resonance energy scales with $T_C$ for different cuprate materials exhibiting a wide range of superconducting transition temperatures \cite{Hufner}, providing tantalizing evidence for a common mechanism related to magnetic fluctuations.

\begin{figure}
\centering\includegraphics[width=1.0\columnwidth]{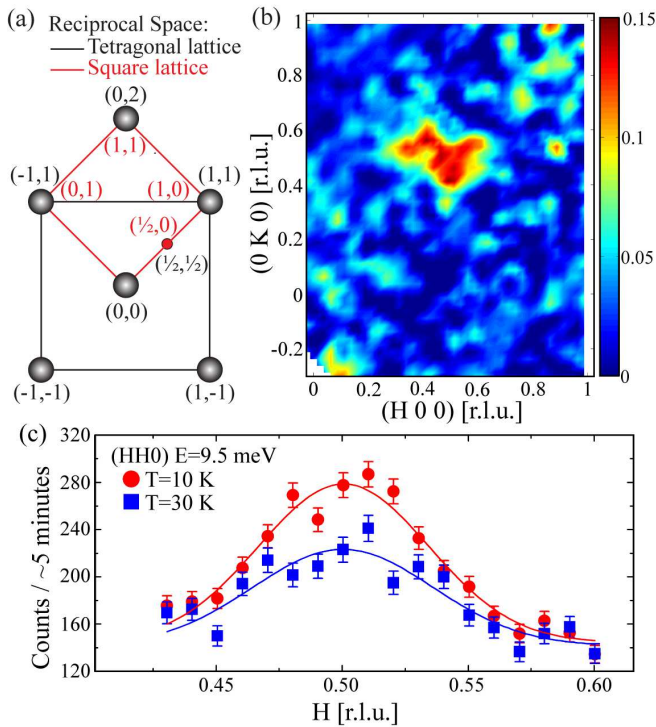}
\caption{\label{fig1} (a) Reciprocal space of both the tetragonal unit cell and the square lattice.  The gray circles indicate the nuclear zone centers.  Labels in black (red) correspond to the tetragonal (square lattice) unit cell.(b) ARCS data collected at T=10 K integrated over the energy range 5 to 25 meV and projected onto the (HK0) plane.  Intensity is shown as a color map.  (c) HB3 constant-E scan along (HH0) at an energy transfer of 9.5 meV for temperatures of 10 K and 30 K.}
\end{figure}

The discovery of a new family of Fe-based high temperature superconductors with $T_C$ as high as 55 K \cite{LaOFFeAsdis,SmOFFeAsdis,CeOFFeAsdis,HPdis,Rotter,Wang,Yeh} presents an exciting opportunity to examine the relationship of spin excitations to the superconducting condensate in unconventional superconductors. The new materials are composed of Fe containing planes (FeAs or FeSe).   Both theory and experiment indicate that simple electron-phonon coupling cannot describe superconductivity in these materials \cite{boeri,christianson1}.
Furthermore, the superconducting state exists in close proximity to magnetism as the parent compounds exhibit spin-density wave order \cite{cruz_lafeaso,mcquire_big}. These observations have been put forth as evidence that the superconductivity in the Fe-based materials is unconventional. The presence of the Fe planes suggests quasi-two-dimensionality, as observed in the cuprates. However, neutron scattering investigations of the spin waves in the undoped parent compounds SrFe$_2$As$_2$ \cite{zhao1}, BaFe$_2$As$_2$ \cite{matan}, and CaFe$_2$As$_2$ \cite{mcqueeney}, indicate anisotropic exchange that cannot be classified as two dimensional. Band structure calculations \cite{singhlda,mazin} indicate that doping should enhance the two-dimensionality of the Fermi surface, favoring superconductivity \cite{mazin}.  Directly probing the magnetic fluctuations in superconducting Fe-based systems is crucial for further progress.

Recent measurements on a polycrystalline sample of Ba$_{0.6}$K$_{0.4}$Fe$_2$As$_2$ found a spin excitation that appears at the onset of superconductivity \cite{christianson2}.   However, it is not possible to unambiguously extract the detailed wavevector dependence of magnetic fluctuations from measurements on polycrystalline samples and, hence, single crystal measurements are essential.

In this letter, we report inelastic neutron scattering measurements on homogeneous single crystals of BaFe$_{1.84}$Co$_{0.16}$As$_2$ (BFCA) with $T_C$=22 K \cite{Sefat3}.  These measurements  indicate a quasi-two-dimensional (2d) magnetic excitation at a wavevector related to the ordering wavevector of the parent compound, BaFe$_2$As$_2$ \cite{Huang,Su}.  The intensity of this excitation is strongly enhanced at the resonance energy upon entering the superconducting state.  The resonance energy is found to be 8.6(0.5) meV or 4.5(0.3) k$_BT_C$, similar to the value observed in the cuprates providing a strong indication that the underlying physics is related at a deep level.

Single crystals of BFCA were grown using FeAs flux.  The starting materials FeAs and CoAs were prepared by heating the elements to 1060 $^{\circ}$C for several hours in silica ampoules followed by furnace cooling to room temperature. BaFe$_2$As$_2$ was prepared by heating Ba and FeAs in a Ta tube to 1200 $^{\circ}$C for 12 h, and 1000 $^{\circ}$C for 36 h, followed by furnace cooling. Ba metal (99.9 \% purity) was handled in a glove box containing less that 1ppm of oxygen and water. All other elements used were better than 99.99 \% pure   For the crystal growth an alumina crucible was loaded with 8 g of BaFe$_2$As$_2$, 1.32 g of CoAs, and 6.54 g of FeAs and sealed in a silica ampoule. The ampoule was heated to 1200 $^{\circ}$C for 24 h and then cooled to 1090 ±C at 1±C/h and removed from the furnace and spun to remove the excess flux.

The resulting crystals exhibit a tetragonal structure with space group I4/$mmm$.  The tetragonal unit cell of BaFe$_2$As$_2$ contains 2 Fe atoms per plane. The Fe atoms in each plane form a square lattice but the near-neighbor Fe-Fe vector lies along a direction rotated by 45$^\circ$ from the tetragonal a-axis (see Fig. 1(a)). With a tetragonal lattice constant, a, the square lattice has a lattice constant of $\tilde{a}$=\emph{a}/$\sqrt{2}$. The wavevector of greatest interest is (1/2 1/2) or equivalently ($\pi$/a,$\pi$/a) in tetragonal notation, equivalent to (1/2,0) or ($\pi/\tilde{a}$,0) in square lattice notation. Electronic structure calculations indicate a Fermi surface nesting instability at this wavector\cite{singhlda,mazin,dong}, and indeed the ordering wavevector in the parent compounds\cite{Huang,Su,zhao2,Goldman} is (1/2 1/2 L) (indexed as (1 0 L) in orthorhombic notation).  As discussed in Ref. \cite{mazin} the different notations are mixed in the literature.  In this letter we use tetragonal notation exclusively.

For the neutron scattering measurements, three single crystals of BFCA with a total mass of 1.8 g were co-aligned in the (HHL) plane.  The data in Fig. 1(b) was collected using the ARCS direct geometry chopper spectrometer at the SNS.  The remainder of the data presented here was collected with the HB-3 triple-axis spectrometer (TAS) at the HFIR configured with collimations of 48$^\prime$-40$^\prime$-80$^\prime$-120$^\prime$ with a fixed final energy of 14.7 meV with pyrolitic graphite (PG) monochromator and analyzer crystals.

\begin{figure}
\centering\includegraphics[width=1.0\columnwidth]{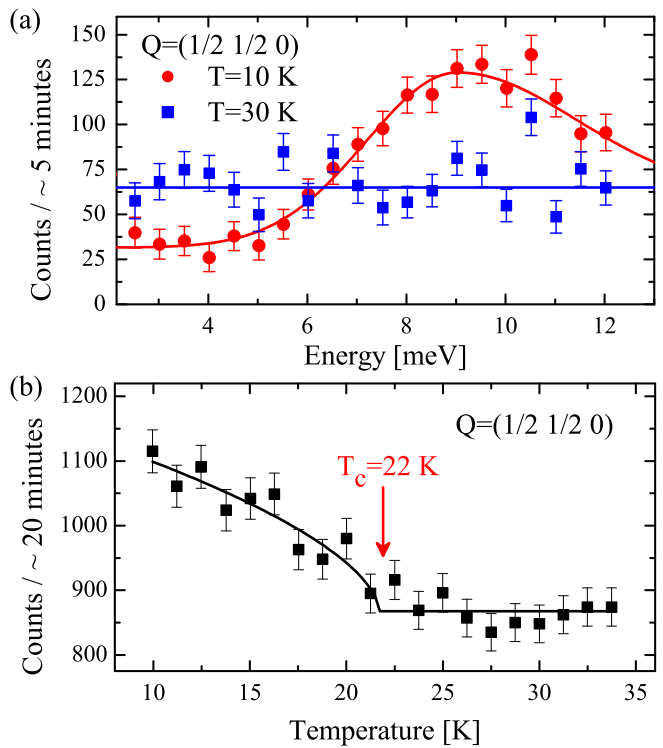}
\caption{\label{fig2} (a) Constant-Q scan, Q=(1/2 1/2 0), for temperatures of 10 K ($T<T_C$) and 30 K ($T>T_C$). (b) Temperature dependence of the inelastic intensity at Q=(1/2 1/2 0) and E=9 meV.  The solid line is a power law fit which results in a $T_C$ of 22(1) K consistent with bulk measurements.}
\end{figure}

Fig. 1(b) shows data collected on ARCS with E$_i$ of 60 meV corrected for empty can background.  The measured data is energy integrated from 5 to 25 meV and projected onto the (HK0) plane with L integrated from -2.5 to 1.5.  This plot shows intensity as a function of Q along (H00) and (0K0).  The data shows a sharp excitation centered at \textbf{Q}=(1/2 1/2).  This observation is confirmed by constant-E scans (E=9.5 meV) along (HH0) measured on the HB-3 TAS (Fig 1(c)) which also show a peak centered at the (1/2 1/2) position.

Fig 1(c) shows scans along (HH0) at temperatures above (T=30 K) and below (T=10 K) the superconducting transition temperature indicating an enhancement of inelastic intensity on passing through $T_C$.  To explore this excess intensity, Fig. 2(a) shows empty can subtracted constant-Q scans with \textbf{Q}=(1/2 1/2 0).  For $T>T_C$, the intensity of the constant-Q scan varies only weakly with energy transfer.  On cooling through $T_C$, the excitation spectrum develops a gap and there is strongly enhanced intensity above the gap energy peaking at a resonance energy of $\sim$9 meV.  This energy is considerably lower than the 14 meV feature observed in Ba$_{0.6}$K$_{0.4}$Fe$_2$As$_2$ \cite{christianson2} suggesting an energy which scales with $T_C$.
It is important to note that the neutron scattering measurements on this Co-doped sample show no evidence for long range antiferromagnetic order, consistent with the published phase diagram \cite{Ning}.

The temperature dependence of the excitation measured at Q=(1/2 1/2 0) and E=9.5 meV is shown in Fig. 2(b). The intensity increases on cooling and a power law fit to the data yields 22$\pm$1 K, consistent with bulk $T_C$ of 22 K.  Clearly, this excitation is strongly coupled with $T_C$ and the resonance energy, $\sim$9 meV, corresponds to a value of 4.75 k$_B$$T_C$ consistent with the canonical value for the cuprates \cite{Hufner} suggesting a remarkable universal behavior between the Fe-based and Cu-based superconductors.

\begin{figure}
\centering\includegraphics[width=1.0\columnwidth]{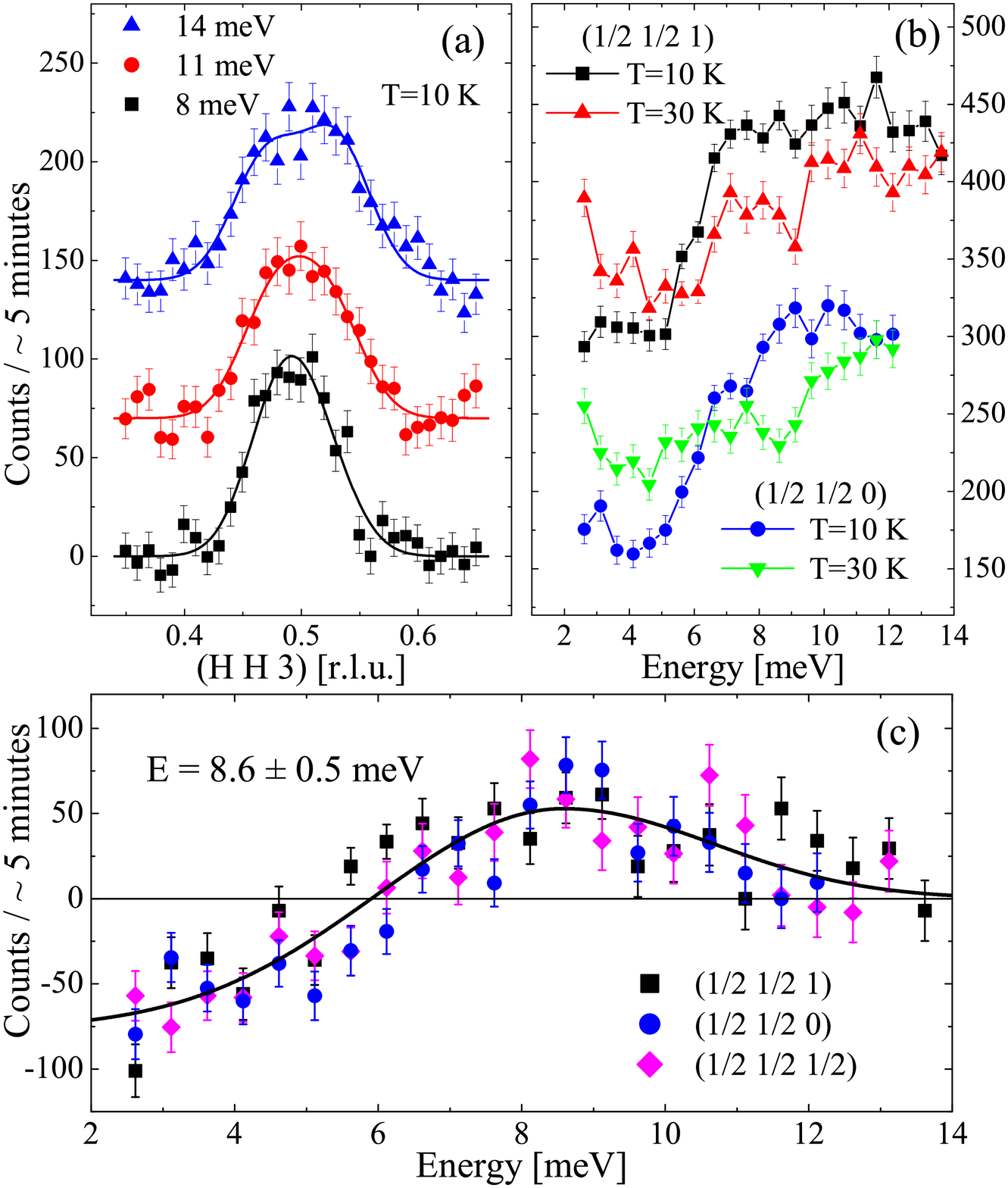}
\caption{\label{fig3} (a) Constant-E scans along (H H 3) for E=8, 11, and 14 meV.  (b) Constant-Q scans for Q=(1/2 1/2 1) and (1/2 1/2 0) for temperatures of 10 K and 30 K.  (c) The 10 K - 30 K difference for constant-Q scans measured at (1/2 1/2 L) with L=0, 1/2, and 1.  The average resonance energy is about 8.6 meV with very weak dispersion.  In panels (a) and (b) the data have been offset for clarity.}
\end{figure}

To examine the in-plane dispersion, constant-E scans were measured near Q=(1/2 1/2 3).  The background subtracted data is shown in Fig. 3(a) for energy transfers of 8, 11, and 14 meV.  These scans show an increase in width with increasing energy transfer consistent with a steeply dispersive excitation.  The best fit to a pair of Gaussians with width constrained to be larger than resolution is represented by the solid lines.  The mode appears to disperse linearly with Q.  To quantify the wavevector dependence, the peak locations were parameterized by a dispersion characterized by a gap together with in-plane and c-axis coupling constants.  Using this dispersion, we estimate an in-plane bandwidth of 70 meV with a lower limit of 60 meV.  To examine the c-axis dispersion, constant-Q scans were performed at (1/2 1/2 L) with L=0, 1/2, and 1 for temperatures of 10 K and 30 K.  We observe the same resonance mode at L=0 and 1 (Fig. 3(b)).  To explore the dispersion of the resonance, Fig. 3(c) shows the difference between the T=10 K and T=30 K constant-Q scans. The data from all three values of L are within statistical error of one another consistent with very weak or no dispersion.  There is, however, a systematic trend as a function of L with the highest peak energy observed at L=0 and the lowest at L=1 with an average of 8.6(0.5) meV.  Our best estimate of the c-axis bandwidth is 0.6 meV with an upper limit of 1.5 meV.  The dimensionality of the spin excitations can be quantified by considering the ratio of in-plane to c-axis bandwidths.  Our estimate of this ratio is 70/0.6=117 with a lower bound of 60/1.5=40.  This result can be directly compared to the spin-wave velocity ratios in the parent compounds where ratios of 2-5 were observed \cite{zhao1,matan,mcqueeney}.  Thus the bandwidth ratio provides quantitative evidence for strongly enhanced two-dimensionality upon doping.

\begin{figure}
\centering\includegraphics[width=1.0\columnwidth]{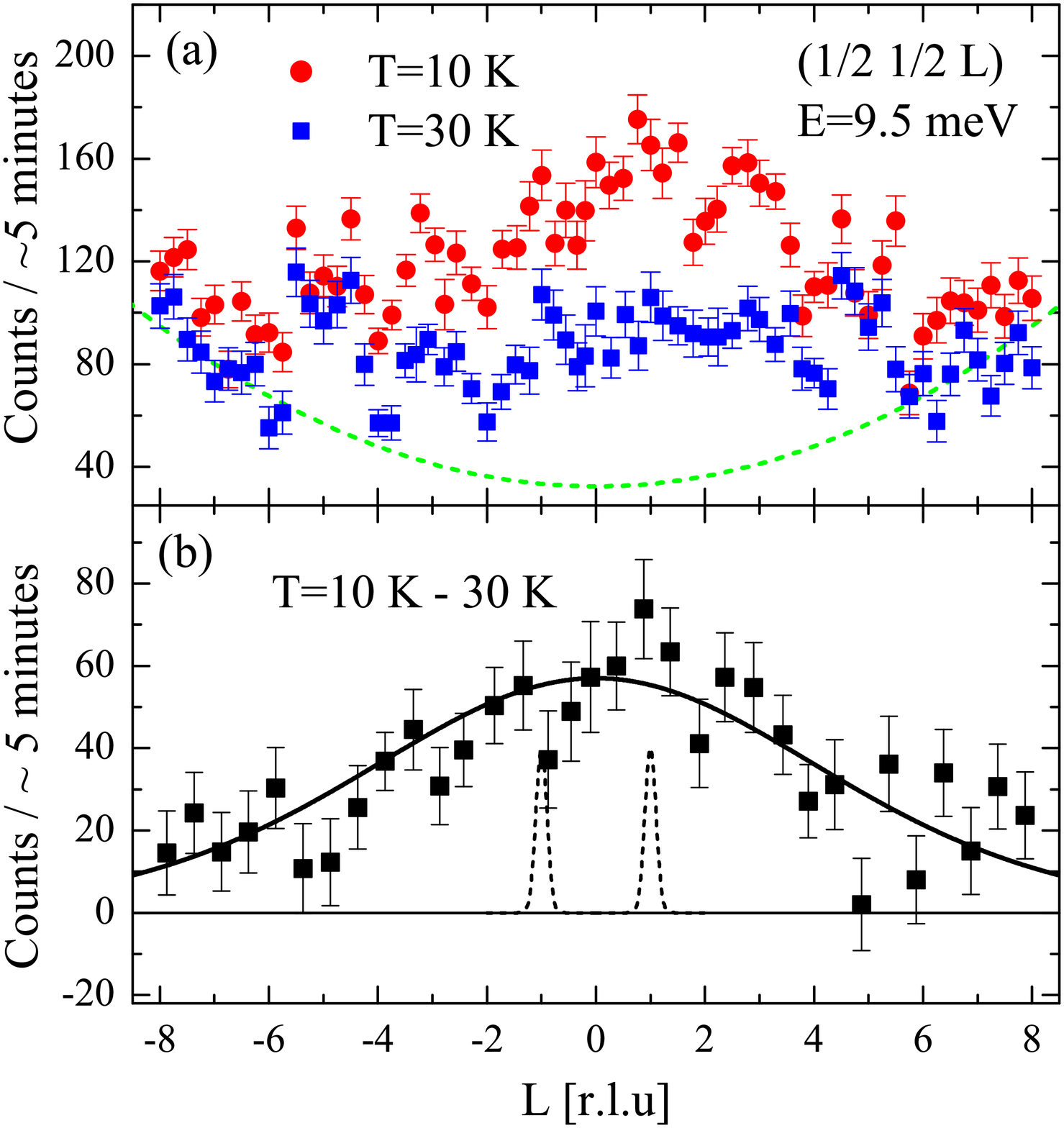}
\caption{\label{fig4} (a) Empty can subtracted constant energy scan with E=9.5 meV along the (1/2 1/2 L) direction at temperatures of 10 K and 30 K.  The dashed line represents an estimate of the non-magnetic background obtained by dividing measurements at 200 K by the Bose factor.  (b) The difference between the 10 K and 30 K scans along (1/2 1/2 L) with E=9.5 meV.  The solid black line is the square of the Fe magnetic form factor. The dashed lines indicate the width of the 10 meV L scan observed in the parent compounds \cite{zhao1,matan,mcqueeney}.  }
\end{figure}

To examine the origin of the excitations, measurements along (1/2 1/2 L) for E=9.5 meV are shown in Fig. 4(a).  The scattering due to the empty sample holder and instrumental background has been removed.  This data shows no obvious periodicity as a function of L.  There is, however, a Q dependent background resulting from multiple and multiphonon scattering as indicated by the estimated non-magnetic background (dashed line in Fig. 4(a)).  Additional understanding of the L dependence can be obtained by subtracting the 30 K data from the 10 K data as shown in Fig. 4(b).  This difference is background independent and the dependence of the intensity on L can be explained entirely by the magnetic form factor.  This is shown in Fig. 4(b) where the solid line is simply the scaled square of the Fe$^{2+}$ spin-only form factor.  This provides strong evidence for the magnetic origin of the scattering.  Moreover, the data is much broader than that measured in the parent compounds, as shown by the dashed lines in Fig. 4(b) taken from constant-E measurements at 10 meV \cite{zhao1,matan,mcqueeney} providing further confirmation of the two-dimensionality of the excitation spectrum.

The form of $\chi^{''}$(\textbf{Q},$\omega$), measured by neutron scattering, has implications for the pairing symmetry of the superconducting state \cite{mazin,Korshunov,maier,christianson2}. Various calculations \cite{symmetry} have found that pairing energetics are very similar for either sign-reversed s-wave, $s_{\pm}$ , or $d_{x^2-y^2}$.  On the other hand, $\chi^{''}$(\textbf{Q},$\omega$) is very different for these scenarios.  Recent calculations of  $\chi^{''}$(\textbf{Q},$\omega$) based on a four band model \cite{Korshunov} find that the $d_{x^2-y^2}$ state produces a very weak resonant peak at an energy less than the superconducting gap, $\Delta_0$.  Conversely, $s_{\pm}$ symmetry leads to $\chi^{''}$(\textbf{Q},$\omega$) strongly peaked at a resonance energy of $\sim$1.4 $\Delta_0$.  Using the measured superconducting gap, $\Delta_0$ = 6.25 meV \cite{Hoffman}, our measured resonance energy of 8.6(0.5) meV corresponds to 1.38(0.08)$\Delta_0$.  Moreover, our data indicates a strong resonance where the intensity at the resonance energy at 10 K is roughly twice that observed at 30 K.  These observations are consistent with the $s_{\pm}$ scenario as predicted in Ref. \cite{Korshunov}.  Finally, the two-dimensional nature of the measured excitation spectrum suggests that 2D models may capture the essential physics of the superconductivity in BFCA.

In summary, inelastic neutron scattering measurements on single crystal samples of BFCA indicate the presence of a resonance in the magnetic excitation spectrum.  The intensity of the resonance, located at the tetragonal (1/2 1/2 L) wavevector, is strongly coupled to $T_C$ and exhibits an average energy of about 8.6 meV = 4.5 k$_B$$T_C$ = 1.38 $\Delta_0$.  The dispersion and L dependence of the excitation spectrum indicates enhanced two-dimensionality in this Co-doped system.

This work was supported by the Scientific User Facilities Division and the Division of Materials Sciences and Engineering, Office of Basic Energy Sciences, DOE. Work at UT was supported by the DOE EPSCoR Implementation award, DE-FG02-08ER46528.

\end{document}